\documentstyle[12pt]{article}
\title{Strings, T-duality breaking, and nonlocality without the
shortest distance}
\author{Hrvoje Nikoli\'c \\
Theoretical Physics Division, Rudjer Bo\v{s}kovi\'{c} Institute, \\
P.O.B. 180, HR-10002 Zagreb, Croatia \\
{\normalsize e-mail: hrvoje@thphys.irb.hr} \\
\makebox[1in]{} \\
}
\date{\today}
\begin{document}
\maketitle
\begin{abstract}
T-duality of string theory suggests nonlocality manifested as the 
shortest possible distance.
As an alternative, we suggest a nonlocal formulation of string 
theory that breaks T-duality at the fundamental level and
does not require the shortest possible distance.
Instead, the string has an objective shape in spacetime at all 
length scales, but different parts of the string interact 
in a nonlocal Bohmian manner. 
\end{abstract}
\vspace*{0.5cm}

\section{Introduction}

T-duality \cite{giv,pol,zwie} is an intriguing property 
of string theory.
In its simplest form, T-duality is a remarkable fact that observable 
properties (such as the spectrum of the mass operator) 
of a closed string compactified on a circle with a radius $R$ 
cannot be distinguished from that of a closed string compactified 
on a circle with the dual radius $\alpha'/R$ 
(where $\sqrt{\alpha'}=l_s$ is the 
fundamental length scale of string theory). 
Effectively, $l_s$ turns out to be the shortest possible 
spacial distance, because a shorter string can always be  
reinterpreted as a longer string in the dual theory. 
The existence of the shortest possible distance 
suggests that a more fundamental formulation of string theory
must be explicitly nonlocal at short distances \cite{seib}.
Another suggestion that string theory is fundamentally nonlocal
\cite{seib,hor} 
comes from the AdS/CFT correspondence \cite{mald}, because this
correspondence demonstrates 
a holographic property of strings, i.e., a property that
observable properties of strings on an anti de Sitter space
can be decribed by a theory that 
lives only on the boundary of this space.
However, the nonlocal nature of strings at the fundamental 
level is still far from being completely understood.

To understand the holographic principle and physics 
at the Planck scale at a more fundamental level, 
't Hooft argues \cite{thooft} that quantum mechanics (QM) 
should be formulated in terms of {\em deterministic hidden 
variables}, according to which QM is probabilistic 
only on the phenomenological level, whereas the more
fundamental hidden (i.e., not observable with present technology)
degrees of freedom satisfy deterministic laws of motion.
Since, owing to the Bell theorem \cite{bell}, 
QM does not allow local hidden variables, the hidden variables 
must necessarily be nonlocal. Smolin argued \cite{smol}
that matrix models, which give a nonperturbative formulation of 
string theory, might also be interpreted as nonlocal 
hidden variables. 
The best known and most successful hidden-variable formulation 
of quantum particles and fields is the
{\em Bohmian} interpretation 
\cite{bohm,bohmrep1,bohmrep2,holrep,holbook,durr,nikolfpl1,nikolfpl2}.
A version of the Bohmian interpretation 
of relativistic particles might be indirectly testable 
even with near-future technology \cite{nikolfpl3}.
The Bohmian interpretation has also been proposed as a possible
interpretation of noncommutative QM \cite{barb}.
Using our results on the manifestly 
covariant canonical quantization of fields \cite{nikolepjc,nikolessay}, 
we have recently argued \cite{nikolepjc2} that the natural formulation of  
strings compatible with the world-sheet covariance on the 
quantum level is, indeed, a formulation in terms
of Bohmian deterministic hidden variables. 
All these results suggest that the Bohmian interpretation 
might provide a more fundamental description of string theory.

In this paper we explore some physical consequences
of the Bohmian deterministic formulation of string theory.
We observe that the formulation of strings 
in terms of Bohmian hidden variables breaks  
T-duality at the fundamental level of the hidden variables.
Thus, from this fundametal point of view, 
$l_s$ can no longer be interpreted as the shortest possible
spacial distance. Nevertheless, it does not mean that  
locality of strings restores. Instead, nonlocality reappears 
in another form, as nonlocality typical of any 
hidden-variable completion of QM, required by the Bell theorem. 

\section{T-duality and hidden variables}

Consider a closed bosonic string compactified on a circle 
in the 25th spatial direction.
A string winding $m$ times around the compactified direction satifies
\begin{equation}\label{X}
X^{25}(\sigma +2\pi)=X^{25}(\sigma)+2\pi m R.
\end{equation}
The momentum in this direction is also quantized:
\begin{equation}
P^{25}=\frac{n}{R}.
\end{equation}
The mass of the string turns out to be given by
\begin{equation}\label{mass}
M^2=\frac{n^2}{R^2}+\frac{m^2 R^2}{\alpha'^2}
+\frac{2}{\alpha'}[N+\tilde{N}-2]  ,
\end{equation}
where the term in the square brackets corresponds to
the contributions from the oscillatory degrees of freedom. 
In (\ref{mass}), the transformation $R\rightarrow \alpha'/R$ 
is equivalent to the exchange $n\leftrightarrow m$.
Therefore, the spectrum of $M$ is invariant under
the T-duality transformation.
Clasically, one can always determine whether the
compactification radius is $R$ or $\alpha'/R$, simply 
by watching the shape of the string that satisfies 
(\ref{X}). In contrast, in QM, if one observes $P^{25}$ or 
$M$, which are described by quantum mechanical operators 
that do not commute with $X^{25}$, then one cannot observe 
$X^{25}$, and thus one cannot observe the shape of the string.
According to the orthodox interpretation of QM, such
quantities that cannot be observed are not physical. 
Consequently, as long as the mass is an observable 
physical quantity, the shape of the string may not be physical,
while physical quantities obey T-duality. Thus, physically,
$l_s$ is the shortest possible distance, which implies
nonlocality. Our point (demonstrated by the discussion above, 
but valid also for T-duality of open strings) is that
the assumption of orthodox interpretation of QM is 
the crucial assumption. Conversely,
if quantities unobservable by the 
standard QM rules are still physical at some more fundamental level, 
then, at this more fundamental level, T-duality may not be a universal 
symmetry. In QM, such hypothetical unobservable degrees of freedom 
are referred to as hidden variables. The Bell theorem \cite{bell} 
shows that any hidden variables compatible with the 
statistical predictions of QM must necessarily be nonlocal.
In the quantum theory of particles and fields, nonlocality 
is usually considered unacceptable. Consequently, 
the Bell theorem is usually considered a strong 
argument against the existence of hidden variables.
However, in string theory, nonlocality seems to be unavoidable, 
so nonlocality can no longer be viewed as an argument 
against hidden variables. Quite the contrary, the existence of 
string nonlocalities, which are still far from being completely understood,
may be viewed as an indication that the principles of QM itself should be 
reformulated in a nonlocal manner, perhaps by using a nonlocal 
hidden-variable formulation of QM. The results of \cite{nikolepjc2}
suggest that this reformulation should be in terms of
the Bohmian hidden variables. In this paper we discuss how 
the Bohmian formulation of strings transcendes T-duality
at the fundamental level 
and introduces a different, clearer, form of nonlocality
manifested as a nonlocal communication between different parts of the string.

\section{Bohmian hidden variables for strings}

The formulation of Bohmian interpretation of particles and fields 
\cite{bohm,bohmrep1,bohmrep2,holrep,holbook,durr,nikolfpl1,nikolfpl2}
is usually based on canonical quantization in the Schr\"odinger picture.  
Accordingly, in this paper, we use the same picture, such that 
the target spacetime covariance is manifest, whereas the world-sheet 
covariance is not. (For the canonical quantization of strings 
in which the world-sheet covariance is also manifest, see
\cite{nikolepjc2}.)
For simplicity, we study only bosonic strings, but
fermionic degrees of freedom can also 
be interpreted in a Bohmian deterministic manner \cite{nikolfpl2},
which allows one to generalize the results of this paper to 
superstrings as well. 

The Hamiltonian operator of a bosonic string is given by
\begin{equation}\label{H}
\hat{H} = 
-\int d\sigma \frac{1}{2} \left[ 
\hat{{\cal P}}^{\alpha}(\sigma) \hat{{\cal P}}_{\alpha}(\sigma)
+ \frac{\partial X^{\alpha}(\sigma)}{\partial\sigma}
\frac{\partial X_{\alpha}(\sigma)}{\partial\sigma} \right] ,
\end{equation}
where 
\begin{equation}
\hat{{\cal P}}_{\alpha}(\sigma) =
i\frac{\delta}{\delta X^{\alpha}(\sigma)} ,
\end{equation}
$\alpha=0,1,\cdots,25$ denotes the target spacetime indices,
the signature of the flat spacetime metric is $(+,-,\ldots,-)$,
$\sigma$ is the affine parameter along the string, and
we use units $\hbar=c=\alpha'=1$. 
Our results will not depend on whether the string is 
closed or open, so we do not specify the boundary conditions.
The quantum state of the string $\Psi[X(\sigma)]$ 
(where $X=\{ X^0,\ldots,X^{25}\}$) is a functional 
of the string configuration described by the functions $X^{\alpha}(\sigma)$.
The state satisfies the Hamiltonian constraint
\begin{equation}\label{sch0}
(\hat{H}-a)\Psi=0 ,
\end{equation}
where $a=1$ when normal ordering is chosen.
The quantity $|\Psi[X(\sigma)]|^2$ 
can be interpreted, at least formally, as the 
probability density for the string coordinates to be 
equal to $X^{\alpha}(\sigma)$. 
(The problem of normalization can be solved by bounding the spectrum 
of the possible values of $X^0$ or by fixing $X^0=\tau$.
However, these technical subtleties will not influence our 
main conclusions, so, for simplicity, we ignore them
in the rest of the analysis.) 
Owing to the Hamiltonian 
constraint, this probability density does not depend
on the world-sheet time $\tau$.
By writing
\begin{equation}\label{psiRS}
\Psi=Re^{iS} ,
\end{equation}
where $R$ and $S$ are real functionals, one finds that the complex
equation (\ref{sch0}) is equivalent to a set of two real
equations
\begin{eqnarray}\label{HJq0}
-\displaystyle\int d\sigma \frac{1}{2} \left[ 
\frac{\delta S}{\delta X^{\alpha}(\sigma)}
\frac{\delta S}{\delta X_{\alpha}(\sigma)}
+\frac{\partial X^{\alpha}(\sigma)}{\partial\sigma}             
\frac{\partial X_{\alpha}(\sigma)}{\partial\sigma} \right]  &
\\
+Q+a & =0 , \nonumber 
\end{eqnarray}
\begin{equation}\label{Rq0}
\int d\sigma \frac{\delta}{\delta X^{\alpha}(\sigma)}
\left[ R^2
\frac{\delta S}{\delta X_{\alpha}(\sigma)} \right] =0 ,
\end{equation}
where
\begin{equation}\label{Q}
Q[X(\sigma)]
=\int d\sigma \displaystyle\frac{1}{2R}  \frac{\delta^2 R}
{\delta X^{\alpha}(\sigma) \delta X_{\alpha}(\sigma)} .
\end{equation}  

The classical Hamiltonian constraint reads $H=0$, so 
$a=0$ in the classical limit.
By restoring units in which $\hbar\neq 1$,  
one finds that 
the right-hand side of (\ref{Q}) attains the additional prefactor $\hbar^2$.
This shows that the classical limit corresponds to 
$Q=a=0$. In this limit, (\ref{HJq0}) takes the form 
of the classical Hamilton-Jacobi equation. However, 
the classical Hamilton-Jacobi equation should be supplemented 
by the deterministic equation of motion of the string
\begin{equation}\label{emhj2}
\left.
-\frac{\partial X_{\alpha}(\tau,\sigma)}{\partial\tau}=
\frac{\delta S}{\delta X^{\alpha}(\sigma)} 
\right|_{X(\sigma)=X(\tau,\sigma)} .
\end{equation}
The Bohmian interpretation consists in the assumption 
that (\ref{emhj2}) is valid even in the quantum case 
with $Q\neq 0$. 
According to this interpretation, 
the string has an objective and deterministic evolution
$X^{\alpha}(\tau,\sigma)$, even when this quantity 
is not measured. All quantum uncertainties are an artefact
of the ignorance of the actual initial conditions
$X^{\alpha}(\sigma)$ at some initial time $\tau$.
If in a statistical ensemble of string configurations
each possible string configuration
$X(\sigma)$ has the probability $|\Psi[X(\sigma)]|^2$ 
at some {\em initial} time, then
(\ref{emhj2}) and (\ref{Rq0}) imply that each possible string configuration
$X(\sigma)$ has the probability $|\Psi[X(\sigma)]|^2$ at 
{\em any} time $\tau$. This explains why 
such a deterministic evolution of strings is in agreement 
with the statistical predictions of the orthodox interpretation of
quantum strings. (For more details on the agreement 
between the statistical predictions of the Bohmian and 
the orthodox interpretation, 
we refer the reader to \cite{bohm,bohmrep1,holbook,nikolfpl1}.)
Eq.~(\ref{HJq0}) is interpreted as the quantum Hamilton-Jacobi 
equation. The quantity (\ref{Q})  
is referred to as the {\it quantum potential}.
From (\ref{emhj2}) and (\ref{HJq0}) one can derive the quantum 
equation of motion 
\begin{equation}\label{eom}
\frac{\partial^2 X^{\alpha}(\tau,\sigma)}{\partial\tau^2} -
\frac{\partial^2 X^{\alpha}(\tau,\sigma)}{\partial\sigma^2} =
\left.
\frac{\delta Q}{\delta X_{\alpha}(\sigma)} 
\right|_{X(\sigma)=X(\tau,\sigma)}.
\end{equation} 
The right-hand side represents the quantum modification of the classical 
string equation of motion. This modification represents the 
quantum force given by the functional gradient of the quantum
potential.   

We stress that,
by postulating the new Bohmian equation of motion (\ref{emhj2}),
we do {\em not} modify any of the standard equations of string theory
that define the physical states,
such as Eq.~(\ref{sch0}). In particular, the mass spectrum 
(\ref{mass}) of string 
states is the same as that in the usual formulation of string theory,
which does not require any new verification. 
Therefore, T-duality at the observable level appears in exactly the same
way as that in the usual formulation of string theory.
Nevertheless, the Bohmian interpretation offers a different interpretation
of that standard result.

The string coordinates $X^{\alpha}(\tau,\sigma)$ are the hidden variables.
Their existence means that the string has a definite position and shape
at all length scales 
even when it is not measured. These hidden variables are not invariant 
under a T-duality transformation. Consequently, T-duality is broken 
at the fundamental hidden-variable level and 
$l_s$ is not the minimal possible length. In fact, at the kinematic 
level, such a quantum string does not differ from the classical one.
However, it does not mean that locality of classical strings is
also restored at the quantum level. Instead, as seen from (\ref{eom}),
nonlocality reappears at the dynamic level. Namely, 
at a given time $\tau$, the quantum force
on the string at the point $\sigma$, in general, 
is not merely a function of $X^{\alpha}(\sigma)$, 
but a functional of the {\em whole} functions $X^{\alpha}(\sigma')$  
at {\em all} points $\sigma'$.
In other words, different parts of the string may communicate
in a nonlocal manner.
To better understand the physical origin of this nonlocality, consider 
a state $\Psi$ that has a form of a local product
\begin{equation}\label{l1}
\Psi[X(\sigma)]=\prod_{\sigma} \psi_{\sigma}(X(\sigma)) .
\end{equation}
In this case, $R$ also takes a similar local-product form,
so (\ref{Q}) implies that the quantum potential is an integral
of the form
\begin{equation}\label{l2}
Q[X(\sigma)]=\int d\sigma\, {\cal Q}(X(\sigma)) ,
\end{equation}
where ${\cal Q}(X(\sigma))$ is a local quantum-potential density.
In this case, the quantum force on the string at the point
$\sigma$ is given by
\begin{equation}\label{l3}
\frac{\delta Q}{\delta X_{\alpha}(\sigma)} =
\frac{\partial {\cal Q}(X(\sigma))}{\partial X_{\alpha}(\sigma)} \equiv
{\cal F}^{\alpha}(X(\sigma)) .
\end{equation}
We see that this force is local, as it depends only on 
$X(\sigma)$. However, in general, the quantum state 
does {\em not} have a local-product form (\ref{l1}). 
(Even the ground state does not have such a form.)
Consequently, unlike (\ref{l2}), the quantum potential cannot be written 
as an integral over a local quantum-potential density ${\cal Q}(X(\sigma))$,
which implies that, unlike (\ref{l3}), the quantum force 
on the string at the point $\sigma$
does not depend only on $X(\sigma)$.
As states that do not have a local-product form (\ref{l1}) exhibit
quantum entanglement, 
we see that
the nonlocal communication between different parts of the string
is a direct consequence of 
quantum entanglement between them. 
Since this nonlocality is realized only on the dynamic level, while 
kinematics is still local and classical, this form of 
nonlocality does not require a radical revision 
of the concept of continuous geometry.  
Nevertheless, by construction, these nonlocal interactions have exactly
the form needed for different parts of the string to conspire so that 
the observable (i.e., not hidden) properties of the string 
obey T-duality.

From a practical point of view, it if fair to note that
the Bohmian hidden-variable formulation of strings does not directly lead
to new observable effects in perturbative string theory.
Nevertheless, we believe that such a formulation
might radically modify the existing attempts to find a
satisfying
{\em nonperturbative} formulation of string theory, as the
existing attempts often significantly deviate from the original picture
of strings as world-sheets in spacetime, whereas the 
Bohmian interpretation suggests that such a picture 
should be taken seriously.

To summarize, the Bohmian hidden-variable formulation of strings, 
suggested by the results of \cite{nikolepjc2}, breaks 
T-duality at the fundamental level and introduces a new form 
of nonlocality into string theory, manifested as 
a nonlocal interaction between different parts of the string.
Such a form of nonlocality does not demand a modification 
of the usual picture of continuous geometry.
\\

\noindent
{\it Acknowledgements.}
This work was supported by the Ministry of Science and Technology of the
Republic of Croatia.


\begin{thebibliography}{99}

\bibitem{giv}
A. Giveon, M. Porrati, E. Rabinovici, Phys. Rep. {\bf 244}, 77 (1994)
\bibitem{pol}
J. Polchinski, hep-th/9611050
\bibitem{zwie}  
B. Zwiebach, A First Course in String Theory
(Cambridge University Press, Cambridge 2004)
\bibitem{seib}
N. Seiberg, hep-th/0601234
\bibitem{hor}
G.T. Horowitz, New J. Phys. {\bf 7}, 201 (2005) 
\bibitem{mald}
J.M. Maldacena, Adv. Theor. Math. Phys. {\bf 2}, 231 (1998) 
\bibitem{thooft}
G. 't Hooft, Class. Quant. Grav. {\bf 16}, 3263 (1999);
hep-th/0003005; hep-th/0104219
\bibitem{bell}
J.S. Bell, Speakable and Unspeakable in Quantum Mechanics
(Cambridge University Press, Cambridge 1987)
\bibitem{smol}
L.~Smolin, hep-th/0201031.
\bibitem{bohm}  
D.~Bohm, Phys.~Rev.~{\bf 85}, 166, 180 (1952)
\bibitem{bohmrep1}
D.~Bohm, B.J.~Hiley,
Phys.~Rep.~{\bf 144}, 323 (1987) 
\bibitem{bohmrep2}
D.~Bohm, B.J.~Hiley, P.N.~Kaloyerou,
Phys. Rep. {\bf 144}, 349 (1987)
\bibitem{holrep}
P.R.~Holland, Phys.~Rep.~{\bf 224}, 95 (1993) 
\bibitem{holbook}
P.R.~Holland, The Quantum Theory of Motion
(Cambridge University Press, Cambridge 1993)
\bibitem{durr}
D.~D\"urr, S.~Goldstein, R.~Tumulka, N.~Zangh\`i, 
Phys. Rev. Lett. {\bf 93}, 090402 (2004) 
\bibitem{nikolfpl1}
H. Nikoli\'c, Found. Phys. Lett. {\bf 17}, 363 (2004)
\bibitem{nikolfpl2}
H. Nikoli\'c, Found. Phys. Lett. {\bf 18}, 123 (2005) 
\bibitem{nikolfpl3} 
H. Nikoli\'c, Found. Phys. Lett. {\bf 18}, 549 (2005) 
\bibitem{barb}
G.D.~Barbosa, N.~Pinto-Neto, Phys.~Rev.~D {\bf 69}, 065014 (2004)
\bibitem{nikolepjc}
H. Nikoli\'c, Eur. Phys. J. C {\bf 42}, 365 (2005) 
\bibitem{nikolessay}
H. Nikoli\'c, hep-th/0601027,  
Honorable Mention of the Gravity Research Foundation 2006 Essay Competition,
to appear in Int. J. Mod. Phys. D.
\bibitem{nikolepjc2}
H. Nikoli\'c, Eur. Phys. J. C {\bf 47}, 525 (2006) 


\end{thebibliography}
\end{document}